\documentclass[twocolumn]{aastex62}

\usepackage{hyperref}

\shortauthors{Batalha et al}

\begin{document}

\title{Color Classification of Extrasolar Giant Planets:  Prospects and Cautions}

\correspondingauthor{Natasha E. Batalha}
\email{batalha@stsci.edu, natasha.e.batalha@gmail.com}

\author[0000-0002-0786-7307]{Natasha E. Batalha}
\affiliation{Space Telescope Science Institute, Baltimore, MD 21218, USA}

\author{Adam J. R. W. Smith}
\affiliation{Department of Astronomy \& Astrophysics, University of California Santa Cruz, Santa Cruz, CA 95064, USA}

\author{Nikole K. Lewis}
\affiliation{Space Telescope Science Institute, Baltimore, MD 21218, USA}
\affiliation{Department of Earth and Planetary Sciences, Johns Hopkins University, Baltimore, MD, USA}
\affiliation{Department of Astronomy and Carl Sagan Institute, Cornell University, 122 Sciences Drive, 14853, Ithaca, NY, USA}

\author{Mark S. Marley}
\affiliation{NASA Ames Research Center, Moffett Field, CA 94035}

\author{Jonathan J. Fortney}
\affiliation{Department of Astronomy \& Astrophysics, University of California Santa Cruz, Santa Cruz, CA 95064, USA}

\author{Bruce Macintosh}
\affiliation{Department of Physics,  Stanford University, Stanford, CA 94305,USA}


\begin{abstract}
Atmospheric characterization of directly imaged planets has thus far been limited to ground-based observations of young, self-luminous, Jovian planets. Near-term space- and ground- based facilities like \emph{WFIRST} and ELTs will be able to directly image mature Jovian planets in reflected light,  a critical step in support of future facilities that aim to directly image terrestrial planets in reflected light (e.g. HabEx, LUVOIR). These future facilities are considering the use of photometry to classify planets. Here, we investigate the intricacies of using colors to classify gas-giant planets by analyzing a grid of 9,120 theoretical reflected light spectra spread across different metallicities, pressure-temperature profiles, cloud properties, and phase angles. We determine how correlated these planet parameters are with the colors in the \emph{WFIRST} photometric bins and other photometric bins proposed in the literature. Then we outline under what conditions giant planet populations can be classified using several supervised multivariate classification algorithms. We find that giant planets imaged in reflected light can be classified by metallicity with an accuracy of $>$90\% if they are \emph{a prior} known to not have significant cloud coverage in the visible part of the atmosphere, and at least 3 filter observations are available. If the presence of clouds is not known \emph{a priori}, directly imaged planets can be more accurately classified by their cloud properties, as oppposed to metallicity or temperature. Furthermore, we are able to distinguish between cloudy and cloud-free populations with $>$90\% accuracy with 3 filter observations. Our statistical pipeline is available on GitHub and can be extended to optimize science yield of future mission concepts. 
\end{abstract}


\section{Introduction}
Atmospheric characterization of directly imaged planets has thus far been limited to ground-based observations of young, self-luminous, Jovian planets \citep{konopacky2013detection, janson2013direct, kuzuhara2013direct,macintosh2015discovery,barman2015simultaneous}. However, near-term space- and ground-based facilities such as the Wide-Field Infrared Survey Telescope (WFIRST, \citet{spergel2013wide}), and the ELTs \citep[e.g.][]{kasper2010epics} along with future concept direct imaging missions (HabEx, \citet{mennesson2016habitable}; LUVOIR, \citet{bolcar2015technology}), could enable the detection of reflected light from cooler ($T_{\rm eff} \sim 150$-$300\,\rm K$) exoplanets located several AU from their host star. 

\begin{figure*}[ht]
\begin{center}
 \includegraphics[origin=c,width=0.8\linewidth]{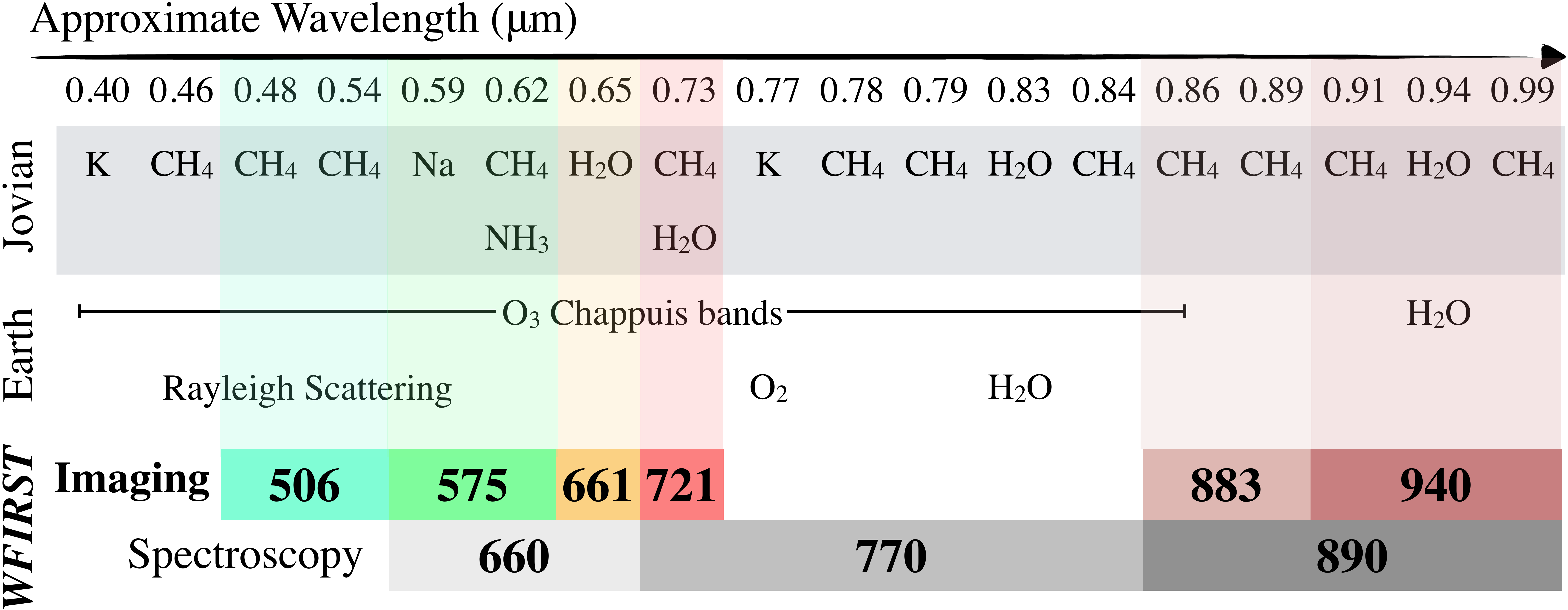}
\caption{Optical absorption features seen in reflected light spectra of gas giant exoplanet models and Earth compared to \emph{WFIRST} observing modes. Absorption features for Jovian systems are taken from \citep{cahoy2010exoplanet} and absorption features for Earth are taken from \citep{krissansen2016is}. \emph{WFIRST} observing modes are labeled by the center of their respective bandpass in units of nm. This work only focuses on the imaging modes. 
\label{fig:wfirst}}
\end{center}
\end{figure*}

The \emph{WFIRST} Corongraph Instrument, CGI,  is currently being refocused as a technology demonstrator. As it is still in the design phase, the 
CGI specifications have not yet been finalized, but the current design does include both photometric and spectroscopic observing modes \citep{noecker2016coronagraph,cady2016demonstration,trauger2016hybrid,seo2016hybrid,balasubramanian2016wfirst}. Since the exact number and wavelength bandpasses for the photometric filters have not been finalized, this work (as shown in Figure \ref{fig:wfirst}) uses the filter set originally defined for the \textit{WFIRST} CGI \citep{spergel2013wide}. Although \textit{WFIRST} may not carry this full filter set, the wavelengths were strategically chosen to explore key molecular features typical of Jovian or Earth-like atmospheres  ($\sim0.45-1.0\,\rm \mu m$) \citep[e.g.][]{des2002remote,cahoy2010exoplanet,macdonald2018exploring}, and they are still representative of the likely capabilities of missions such as HabEx \& LUVOIR \citep{mennesson2016habitable,bolcar2015technology}.

Jupiter-sized planets located a few AU away from their parent stars are particularly interesting objects
for direct imaging in scattered light. They are large and, if cloudy, bright, even at a few AU separation, and thus offer
higher contrast ratios at larger angular separations than habitable zone terrestrial planets. Such giants are not, however, 
Jupiter twins. The cloud decks of planets somewhat warmer than Jupiter will progressively evaporate as first the uppermost $\rm NH_3$ and eventually even the deep $\rm H_2O$ clouds evaporate \citep{marley1999reflected,sudarsky2000albedo,sudarsky2003theoretical}. The nature and detailed characteristics of the uppermost cloud deck define the reflected light continuum \citep{cahoy2010exoplanet,macdonald2018exploring} and the depths of absorption bands. Because the effects of clouds are so important, \citet{sudarsky2000albedo} suggested that cloud type should be used to classify giants, with classes I through III representing the sequence from ammonia to water to no clouds.

To explore how well such planets could be characterized by reflected light spectra of the quality which might be expected from optical wavelength coronagraphic imaging, \citet{lupu2016developing} coupled a reflected light model based on \citet{cahoy2010exoplanet} and an instrument noise model based on \citet{robinson2016characterizing} to a nested sampling algorithm in order to demonstrate the ability to retrieve cloud properties, methane abundance, gravity, and cloud-top pressure. They concluded that the presence of clouds and methane could be determined with high confidence, with clouds being even
easier to identify (from the continuum shape) than the presence or absence of methane. \citet{nayak2017atmospheric} expanded on this work to explore how poorly constrained or unknown planet radii and planet-star phase angles could impact retrievals of high fidelity planet properties. Specifically, they determined that a signal-to-noise ratio (SNR) of 20 was required to retrieve accurate methane abundances at all phase angles.

Most recently, \citet{lacy2018characterization} used an independent reflected light and instrument noise model based on the work of \citet{madhu2012analytic} and \citet{nemati2017sensitivity}, respectively. They determined that for the most favorable known radial velocity targets, only exposure times of hundreds of hours would enable the measurement of planet radius and methane (if the phase functions and orbital parameters were known \emph{a prior}). 

With regard to terrestrial planets, \citet{feng2018characterizing} explored retrievals of atmospheric and planetary properties for Earth twins as a function of SNR and spectral resolving power. As expected from previous work on Jovian systems, they concluded that \emph{WFIRST}, combined with a starshade, would provide limited astrophysical information of Earth-type planets because prominent spectral features, such as water vapor, ozone and oxygen, can only be weakly detected with SNR=20 data. 

The consensus among all of this work, regardless of planet-type, is that it will be challenging and time intensive to attain sufficient spectroscopic SNR on many targets in the era of \emph{WFIRST} and future space-based coronographs \citep{traub2016science}. Therefore, in light of the upcoming 2020 decadal survey, it would behoove the community to determine if additional observing techniques could aid in the characterization of exoplanets or the optimization of target selection. In particular, this sentiment has motivated the idea of leveraging photometric surveys as a means of selecting scientifically interesting targets for time-intensive follow-up, and even as a means of broadly classifying planet-types. The dominant noise source for corongaraph or starshade spectrographic instruments on small or medium telescopes such as \textit{WFIRST} is often detector dark current \citep{nemati2017sensitivity}. As a result, exposure times scale with the number of pixels a spatial and spectral resolution element occupies - 16 or more pixels for a typical integral field spectrograph. In narrow-band photometric imaging, the signal from a single absorption feature occupies only four pixels. 

Because of how stellar photometry was leveraged to create the Hertzsprung-Russell diagram, \citet{traub2003extrasolar} reasoned that the physical characteristics of a planet and its atmosphere could be derived from its colors. This work was based on an analysis of the 0.3-1 $\mu$m reflected light spectra of the Solar System planets, the Moon and Titan, all at full phase.  However \citet{cahoy2010exoplanet} pointed out that the color of planets changes with phase and thus care would have to be taken in identifying planets based solely on color. \citet{krissansen2016is} conducted a comprehensive analysis of Solar System bodies and modeled exoplanets, focusing specifically on separating a habitable Earth-twin spectrum from an uninhabitable planet. Out of their sample of spectra, the Earth’s reflectance spectrum was the only one to have a ``U-shape’’, caused by biogenic O$_2$. Therefore, they concluded that photometry, as a preliminary characterization tool, could be valuable.

However, the brightness in any given bandpass of a planet is a complex function of several parameters. Exoplanets will likely exhibit a great diversity of atmospheric properties not fully encompassed by the Solar System sample. The most dominant parameters that affect the brightness profile are 1) the abundances of absorbing gasses, which can be studied via the metallicity, 2) the cloud height and scattering properties, which largely affect the brightness in the near infrared, 3) photochemical haze abundances and their optical properties, 4) the relative importance of scattering versus absorption, which also depends on the cloud and haze optical depths, 5) the gravity and temperature, which will impact the scale height, 6) the phase angle, which determines how important forward and back scattering is, and 7) the bandpass of the observing filter.

With such a large parameter space and so many confounding variables, it is difficult to analyze all of these parameters with computationally intensive retrievals methods \citep[e.g.][]{lupu2016developing}. Yet, in the era of next-generation direct imaging missions, it will be important to numerically assess the influences of each of these parameters on our ability to yield scientifically interesting results. 

Therefore, as opposed to using retrieval algorithms,  we assess the intricacies of these effects by analyzing a grid of 9,120 reflected light spectra spread across metallicity, pressure-temperature profiles, cloud properties, and phase angles. This large dataset allows us to leverage several multivariate classification algorithms, which are computationally effective, and give numerical insights into our ability to classify giant planets by physically motivated groups using photometric observations. In doing so, we answer the following questions: 
\begin{enumerate}
    \item Is there are a strong correlation between atmospheric properties and \emph{WFIRST}-like optical filters?
    \item If so, can those correlations be leveraged to create color-color diagrams that separate planets into any physically-motivated groups?
    \item If not, do any more sophisticated ``machine learning'' algorithms succeed in correctly classifying planets given our theoretical training set? 
    \item And, are there certain band passes (or a minimum number of band passes) required for successful classification? 
\end{enumerate}
Section \ref{sec:model} describes our forward modeling methodology for creating pressure-temperature profiles, chemical profiles, clouds, and reflected light spectra. Section \ref{sec:stats} describes our statistical methodology for classifying planet types. Section \ref{sec:results} contains our results, and lastly, our discussion and concluding remarks are in Section \ref{sec:disc} \& \ref{sec:conc}.

\section{Modeling Approach} \label{sec:model}
There are three components required for the computation of reflected light spectra: 1) the temperature-pressure ($T(P)$) profile, 2) the composition of the atmosphere as a function of altitude, and 3) the scattering properties of the clouds. We vary each of these components independently from each other, neglecting any feedback between cloud production and atmospheric temperature. Additionally, all our models are run for a single gravity planet ($25\,\rm m\, s^{-2}$) that orbits a Sun-like star. In Section \ref{sec:disc} we discuss the impact of expanding our analysis to incorporate additional gravities and stellar types. 
\subsection{The Temperature \& Chemical Structure}
Our $T(P)$ profile is computed using the radiative-convective model initially developed by \citet{mckay1989thermal} and modified for irradiated gas giant planets in \citet{marley1999thermal,marley2002clouds,fortney2005comparative,fortney2008unified}. It has also been used in many other comparisons with data \citep[e.g.][]{marley1996atmospheric,roellig2004spitzer,mainzer2007moderate, saumon2008evolution}. 
The model requires, as input, an internal heat flux and the incident flux from the parent star. We keep the internal heat flux equivalent to 150 K (similar to that of Jupiter \citep{fortney2007planetary}) for all the cases considered here. For incident flux, we use semi-major axis as a proxy and explore 10 different cases, ranging from 0.5-5.0 AU from a Sun-like star. Throughout the text we refer to planet-star distance as ``distance'', and it can be assumed this also indicates a variation in temperature-pressure profile. The model then iterates to radiative-convective equilibrium using the two-stream source function technique of \citet{toon1989rapid}. 

\begin{figure}[ht]
\begin{center}
 \includegraphics[origin=c,width=0.9\linewidth]{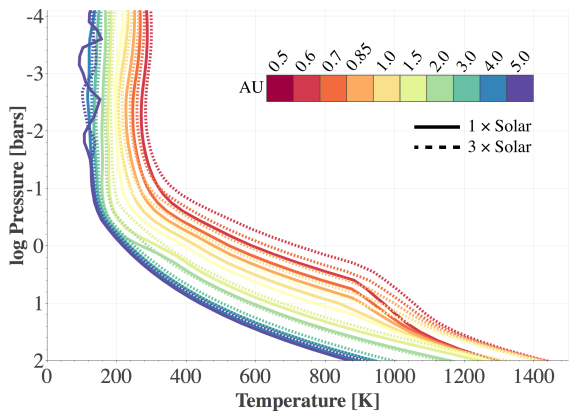}
\caption{Representative pressure-temperature profiles used as input to model the albedo spectra. All cases are for a planet with a gravity of 25 m s$^-2$ around a Sun-like star. \textbf{Main point}: Our grid covers a wide range of parameter space in temperature, which allows us to explore a diversity of chemical compositions and cloud condensates.
\label{fig:tp}}
\end{center}
\end{figure}
We compute the composition of the atmosphere by using the chemical equilibrium grid most recently updated in \citet{lodders2006chemistry}. We consider 6 different metallicities ranging from 1-100$\times$~Solar using the initial elemental abundances from \citet{lodders2003solar}. The optically relevant gases are H$_2$, H, CO$_2$, CO, CH$_4$, NH$_3$, N$_2$, H$_2$S, Na, K, TiO \& VO. This  also  includes  continuum opacity  from  H$_2$-H$_2$,  H$_2$-He,  H$_2$-H,  and  H$_2$-CH$_4$. The opacities for these gases were calculated in \citet{freedmn2008line} and updated in \citet{freedman2014gaseous}. 

\citet{macdonald2018exploring} created a similar grid of parameterized $T(P)$ profiles by fitting to the self-consistent model used in this work. We verified that the $T(P)$ profiles derived in this work are in agreement with the parameterized profiles from \citet{macdonald2018exploring}.

Figure \ref{fig:tp} shows several representative $T(P)$ profiles for 1$\times$ and 3$\times$~Solar metallicities for the full range of semi-major axes explored here. We emphasize that all our models are computed at a gravity of $25\,\rm m\,s^{-2}$ (gravity of Jupiter). This serves as an initial staring point for assessing the use of colors in classifying planets. In Section \ref{sec:disc} we discuss how our results may be further impacted by different gravities/unconstrained masses.

\subsection{The Cloud Model}
The full grid, partially seen in Figure \ref{fig:tp}, contains a wide parameter space in temperature. Therefore, depending on the case, we must consider the condensation of different cloud species. On the cooler end, toward 5 AU, we have the potential formation of H$_2$O clouds \citep{marley1999reflected,sudarsky2000albedo, morley2012neglected}. As we go towards hotter systems we must consider the potential condensation of several other species: NH$_3$, KCl, ZnS, Na$_2$S, MnS, Cr, MgSiO$_3$ also discussed in \citet{morley2012neglected}.   

We calculate the cloud properties using a Mie scattering treatment of particle sizes calculated from a widely-used cloud model \citet{ackerman2001precipitating}. Specifically, the cloud mass and particle sizes are calculated by solving 
\begin{equation}
    K_{zz} \frac{\partial q_t}{\partial z} + f_{\rm sed}wq_c = 0
\end{equation}
where $K_{zz}$ is the vertical eddy diffusion coefficient ($\rm cm^2\,s^{-1}$), $q_t$ is the volume mixing ratios of the total condensing species (condensate plus vapor form), $q_c$ is the volume mixing ratio of just the condensate, and $w$ is the convective velocity. Lastly, $f_{\rm sed}$ is a parameter generally used to tune the sedimentation efficiency of the atmosphere. High values of $f_{\rm sed}$ (i.e., $>$1) create thin cloud profiles with large particles, which are generally optically thin. Low values of $f_{\rm sed}$ (i.e., $<$1), result in the opposite-- thick cloud profiles with small particles, which are generally optically thick. For reference, reflected light observations of Jupiter are well-fit with $f_{\rm sed}=3$ \citep{ackerman2001precipitating,cahoy2010exoplanet}. For each $T(P)$-metallicity point, we compute cloud profiles for $f_{\rm sed}=0.01-6$, where lowest bound $f_{\rm sed}$ values are motivated by observations of hot Jovian worlds \citep{demory2013inference}.

Ultimately, the cloud model calculates the wavelength-dependent optical depth, the single scattering albedo and the scattering asymmetry factor, which are the last components needed for the the reflected light calculation. 

\subsection{Reflected Light: The Albedo Model}
The wavelength-dependent \emph{geometric} albedo is the ratio of reflected planetary flux at full phase to the flux from a perfect Lambert disk. Realistically, direct imaging observations of exoplanets will never yield full phase observations because full phase occurs too spatially near the parent star. Our albedo model was originally constructed for full phase Titan observations \citep{mckay1989thermal}, but has since gone through substantial developments. \citet{marley1999thermal} and \citet{marley1999reflected} extended the model to giant planets (both Solar System and exoplanets). Later, \citet{cahoy2010exoplanet} added the ability to compute phase-dependent albedo spectra, crucial for the study of directly imaged planets. The radiative transfer of the model closely resembles our $T(P)$ model (since the models both originate from \citet{mckay1989thermal}). Therefore, the radiative transfer calculations use the same \citet{toon1989rapid} scheme described in Section \ref{sec:model}.3. Likewise, the opacities considered here are identical to those in the $T(P)$ model. 
We feed the results of each $T(P)$-metallicity-cloud profile to compute albedo spectra for phases between 0-180 degrees in increments of 20 degrees. This completes the last component of the full albedo grid.
\begin{figure*}[ht]
\centering
 \includegraphics[origin=c,width=0.7\linewidth]{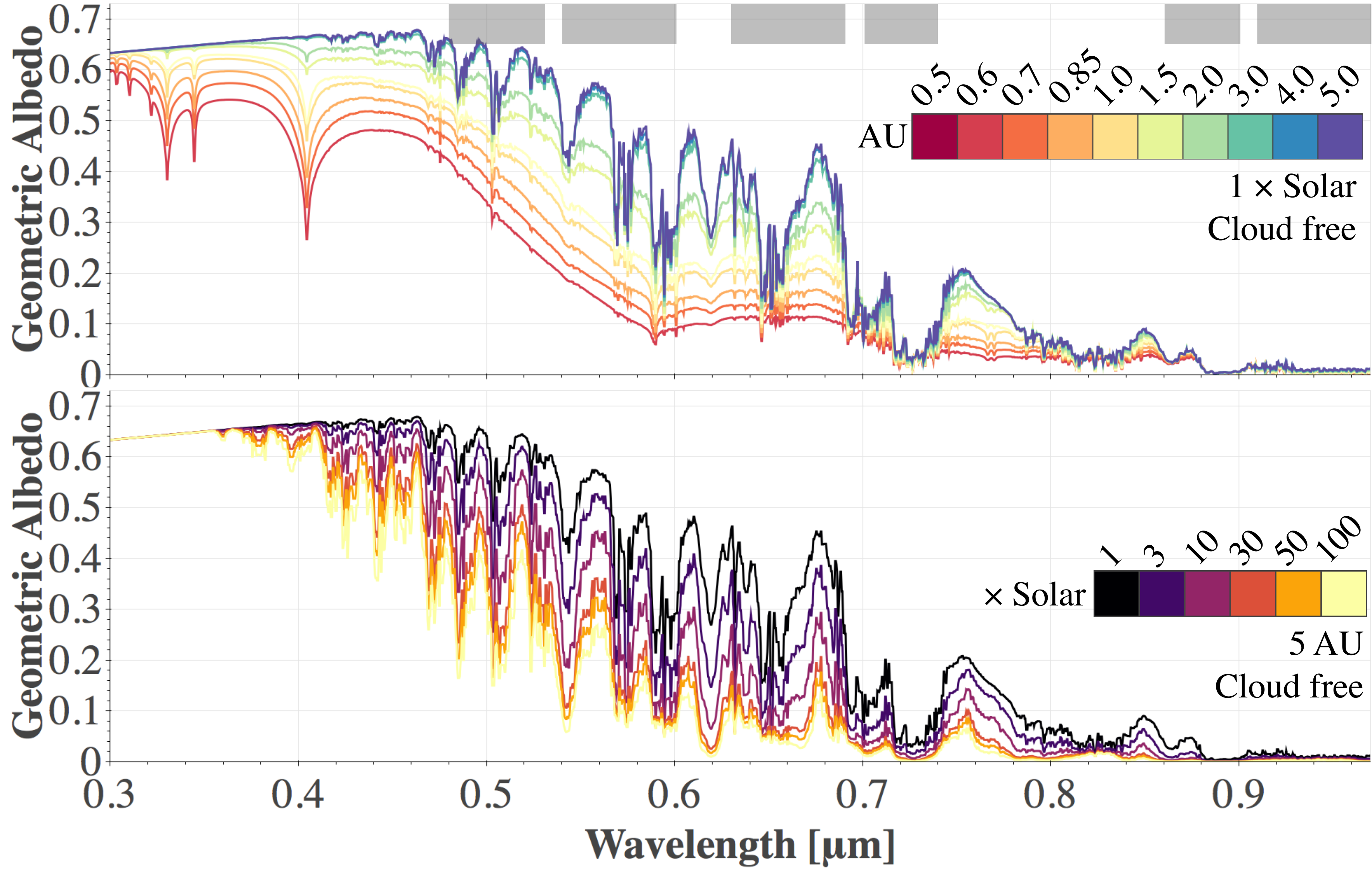}
\caption{Representative albedo spectra showing the 10 different parent-star distances, the 6 different metallicities, and the 6 different bandpasses explored. The top panel shows the cases for a single metallicity (1$\times$~Solar) with all available distances from host star. The bottom panel show cases for a system with a parent-star separation of 5 AU, with all available metallicities. \textbf{Main Point}: Temperature dictates the main opacity source (Rayleigh and alkali dominate as temperature increases, and CH$_4$ \& H$_2$O dominate as temperature decreases). Metallicity dictates the total overall opacity of the atmosphere (atmosphere becomes darker for higher metal content). 
\label{fig:meh+au}}
\end{figure*}

\begin{figure}[ht]
\begin{center}
 \includegraphics[origin=c,width=\linewidth]{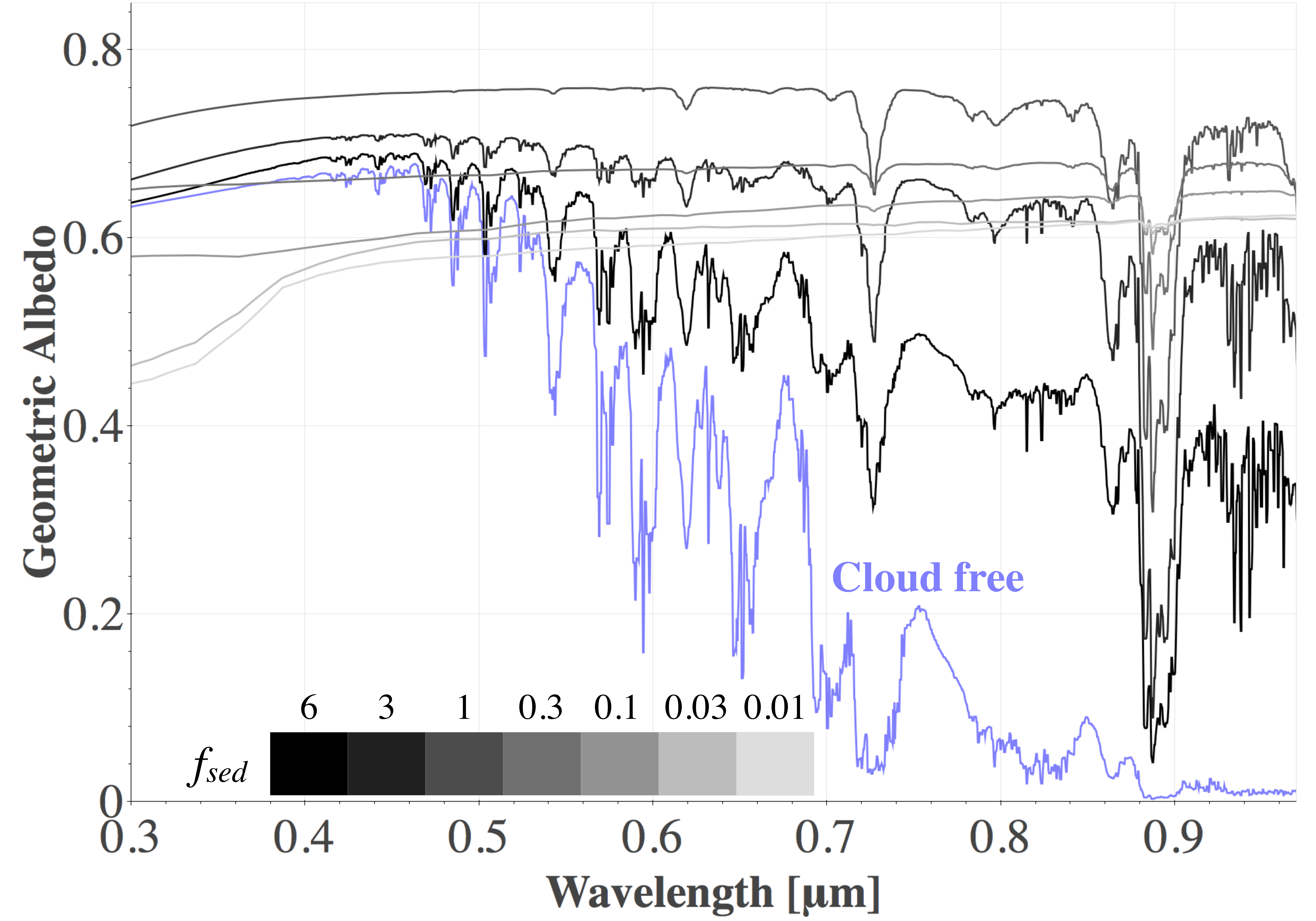}
\caption{Representative albedo spectra showing the effect of varying cloud profiles for a 1$\times$~Solar composition planet located 5 AU from a Sun-like star (gravity=25m/s$^2$). Our cloud profiles are varied by increasing values of $f_{\rm sed}$, the sedimentation efficiency. \textbf{Main Point}: 1) Large $f_{\rm sed}$'s create vertically thin, optically thin clouds and vice versa, 2) Clouds increase atmospheric brightness toward 1 $\mu$m. 
\label{fig:cloud}}
\end{center}
\end{figure}

\begin{figure}[ht]
\begin{center}
 \includegraphics[origin=c,width=\linewidth]{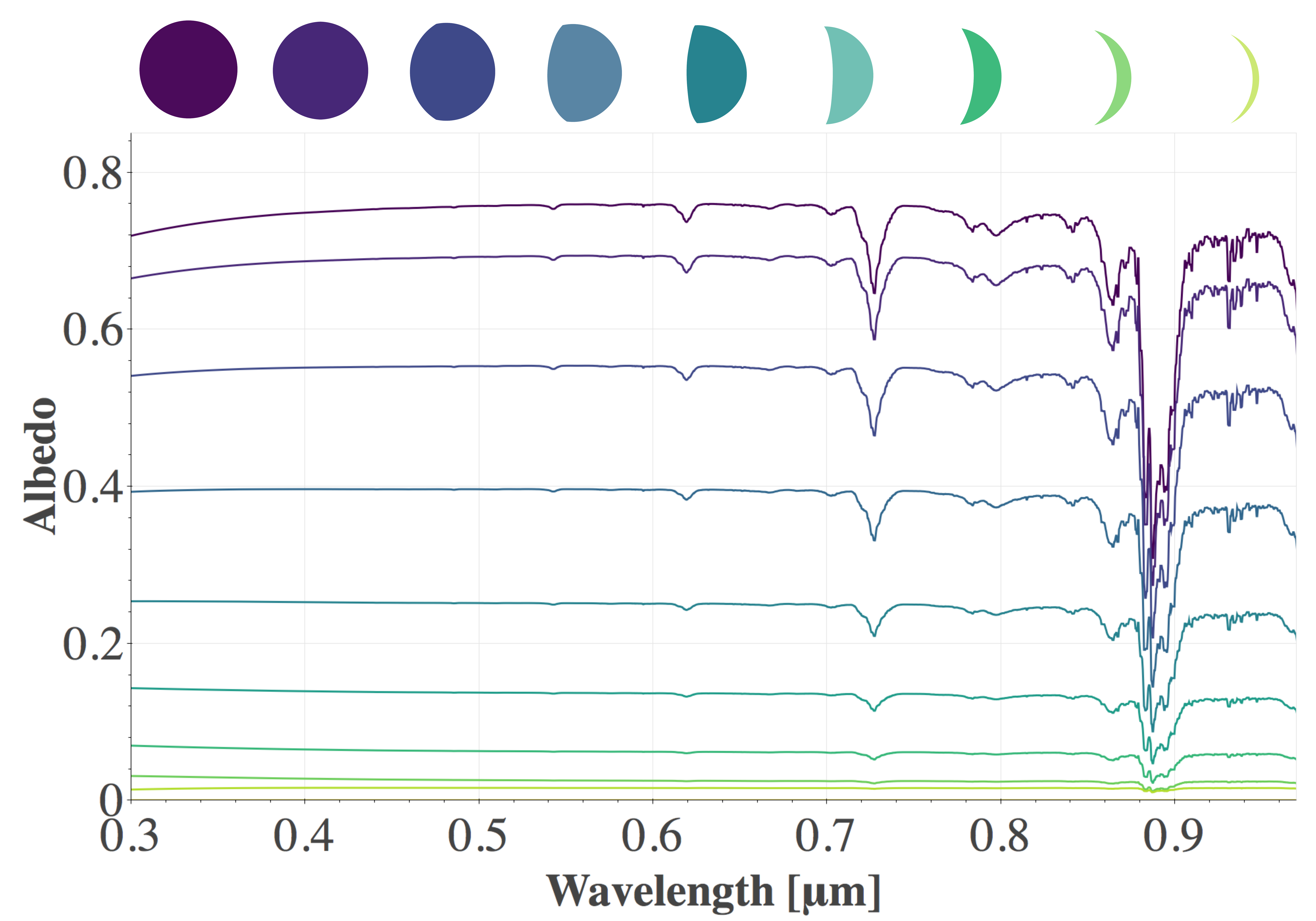}
\caption{Representative albedo spectra showing the effect of phase when clouds are also present. All models are for a 1$\times$~Solar Jupiter-analogue located 5 AU from a Sun-like star with a cloud profile with $f_{\rm sed}$=3. \textbf{Main Point}: Higher phase (from full phase=0) observations decrease the overall brightness of the directly imaged planet. 
\label{fig:cloud+phase}}
\end{center}
\end{figure}
The top panel in Figure \ref{fig:meh+au} shows the effect of decreasing temperature on the geometric albedo for a cloud-free Solar metallicity case. As found by \citet{sudarsky2000albedo} and \citet{cahoy2010exoplanet}, starting at high temperatures ($a_s$=0.5 AU), the spectra of 1$\times$Solar Jovian worlds are dominated by alkali opacities and Rayleigh scattering. The clear atmosphere allows photons to penetrate deep in the atmosphere where temperatures exceed 800~K and Na \& K are in relatively high abundance due to chemical equillibrium. As temperatures decrease (toward $a_s$=2 AU), alkali opacity becomes less dominant because cooler temperatures at chemical equilibrium favor the production of H$_2$O and CH$_4$. Na \& K are lost into sulfide and salt condensates \citep{lodders199alkali,morley2012neglected}. H$_2$O and CH$_4$ prevent photons from penetrating deep into the atmosphere and overall, results in a brighter albedo spectrum. From $a_s$=3-5 AU, there is little observable effect on the spectrum due to relatively minor changes in the $T(P)$ profile (see Figure \ref{fig:tp}). The presence or absence of Na \& K lines short of 0.41 $\mu$m would be strong temperature indicators. However, they are outside the current proposed range of \emph{WFIRST} band passes. 

The bottom panel in Figure \ref{fig:meh+au} shows the effect of increasing metallicity on the geometric abledo for a cloud-free, $a_s$=0.5 AU case. Enhancing metallicity from Solar values increases the total opacity of the atmospheres, which leads to a decrease in the atmospheric brightness (or deeper molecular bands).  The biggest spectral changes occur from 1-10$\times$~Solar, after which, the overall albedo spectra remains relatively unchanged. Ultimately this will lead to challenges in accurately constraining metallicites above 10$\times$~Solar, which we discuss in Section \ref{sec:results}.

Figure \ref{fig:cloud} shows the effect of increasing the sedimentation efficiency, $f_{\rm sed}$, in our cloud model. Lower values of $f_{\rm sed}$ lead to more vertically extended, optically thick clouds, which can be seen in Figure \ref{fig:cloud}. In this 5 AU, Solar metallicity case, as $f_{\rm sed}$ increases, H$_2$O clouds begin to dominate the spectrum and drastically increase the brightness past $\sim0.6$~$\mu$m. This brightness peaks at f$_{sed}=1$ then begins to decease for f$_{sed}<1$, as thick clouds decrease the Rayleigh scattering efficiency. 

Lastly, Figure \ref{fig:cloud+phase} shows the effect of our albedo model's phase dependence, via work done by \citet{cahoy2010exoplanet}. As expected the brightness of the atmosphere decreases with decreasing phase. Not pictured, is the degenerate effect between phase and unknown planet radii, not explored in this work (see \citet{nayak2017atmospheric} for a comprehensive analysis of those degeneracies). 

\section{Statistical Approach}\label{sec:stats}
Our full theoretical population includes: 6 metallicities $\times$ 10 $P(T)$ profiles $\times$ 8 cloud profiles $\times$ 19 phase angles= 9,120 model planetary spectra. We assume that each of these 9,120 planets are observed with the 6 imaging filters depicted in Figure \ref{fig:wfirst}. We compute absolute magnitudes and colors for all possible combinations of filters. 

This number of planets is \emph{of course} grossly over estimating what a mission such as \emph{WFIRST} (or even future mission concepts such as LUVOIR or HabEx) would provide observationally. However, we specifically aim to determine whether or not colors can be used under any scenario to classify planets, as the field of stellar astrophysics has done with stars. 

Under these assumptions, our statistical problem is a classic \emph{supervised classification multivariate analysis} problem \citep{feigelson2012modern}, because we assume that we know the properties of each of the 9,120 planets. The ultimate goal is to determine if any modern algorithms can correctly determine physical planet properties given photometric observations in WFIRST-like filters. This is a fundamentally different approach than that of \citet{krissansen2016is}. In their analysis, the goal was to determine if there were any optical filter combinations that could separate an Earth-twin from every other type of hypothetical exoplanet. Therefore, their algorithm chose filters which maximized distance in color-color space from Earth to their theoretical population. 

There are dozens of open-source algorithms widely used in multivariate classification problems \citep[][and refs. therein]{johnson2002multivariate,izenman2008modern,feigelson2012modern, marsland2015machine}. Although traditionally these algorithms are available through \textbf{R} packages, Python also has a widely used package for machine learning problems called \texttt{scikit-learn} \citep{scikit-learn}, which we adopt here. All of the code used in our statistical analysis is publicly available on \texttt{Github}\footnote{\url{https://github.com/natashabatalha/color-color}} and we footnote specific Python functions used throughout the text, when applicable. 

We first determine the effectiveness of the classification algorithms by using the k-fold\footnote{\texttt{sklearn.model\_selection.KFold()}} cross validation\footnote{\texttt{sklearn.model\_selection.cross\_val\_score()}} method \citep{kohavi1995study}. In this method, for each algorithm we: 1) randomly separated the original sample into $k$ equal groups, 2) keep one of the groups as the validation data for testing, while the other remaining groups are used as the training data, and 3) compute the mean predictive accuracy given the training and validation set. These three steps are then repeated $k$ times so that each group is used at least once as the validation data. In this analysis we also ensure that the same random seed is used to split the data into identical groups each time an algorithm is evaluated. We test our cross validation routine with values of k $k$ from $k=10$-$100$. 

We briefly describe each of the algorithms as they relate to the problem of directly imaged planets, with further technical reading available in \citet{johnson2002multivariate,izenman2008modern,feigelson2012modern, marsland2015machine}: 
\begin{itemize}
    \item \emph{Linear Discriminant Analysis\footnote{\texttt{sklearn.discriminant\_analysis.LinearDiscriminantAnalysis()}} (LDA)} is very similar to a Principal Component Analysis (PCA) in its attempt to find the component axes that maximize the variance of the data. PCA, however, is an \emph{unsupervised algorithm} and therefore ignores class labels (in our problem class labels would be information such as metallicity, temperature, cloud properties). LDA is \emph{supervised}, which means it considers class labels and attempts to project the data into directions that maximize the separation between these classes. Those directions are called the linear discriminants. In Section \ref{sec:results}.3 we will compute linear discriminants and assess whether they are successful in separating out planet populations observed with WFIRST-like photometric filters.
    \item \emph{K-Nearest Neighbors (KNN)}\footnote{\texttt{sklearn.neighbors.KNeighborsClassifier()}} is commonly used when nonlinearities exist between classes in a dataset. For example, if two different low metallicity populations appeared in distinct regions in color-color space (e.g. cloud-free versus cloudy), then a linear method would readily fail. The KNN algorithm uses the training data to compute on-the-fly calculations about the similarity between any given input and the training data. Therefore, KNN would theoretically be able to classify metallicity given varying cloud profiles, if the different cloud profiles created distinct populations of metallicities. We use the grid search\footnote{\texttt{sklearn.grid\_search.GridSearchCV()}} method in \texttt{scikit-learn} to optimize the number of nearest neighbors to use in addition to testing its sensitivity to standard values, such as the Python package's default value. 
    \item \emph{Classification and Regression Trees\footnote{\texttt{sklearn.tree.DecisionTreeClassifier()}} (CART)} recursively splits the data by creating a prediction model at each partition \citep{breiman2017classification}. At each partition, the algorithm aims to minimize the inhomogeneity of the dataset, called impurities. The two methods to evaluate impurities included within \texttt{scikit-learn} are 1) the \emph{Gini} method and 2) the information gain methods. We test both to ensure insensitivity to our result. The result of a CART analysis can be represented graphically in the form of a decision tree. We tune the model using the same method described in KNN. Additionally, we take extra precaution to avoid highly complex and nonintuitive decision trees that over-fit the data (discussed further in Section \ref{sec:results}).  
    \item We also analyzed the results of a Multi-Class Logistic Regression\footnote{\texttt{sklearn.linear\_model.LogisticRegression()}}, and a naive Bayes classifier\footnote{\texttt{sklearn.naive\_bayes.GaussianNB()}}, and a Support Vector Machines\footnote{\texttt{sklearn.svm.SVC()}} but do not present the results because the models failed to classify planets by physical properties under all scenarios with a success rate of $\le1$\%. 
    
\end{itemize}
We test each algorithm under different observing scenarios: if 2 filters were available, 3 filters, 4 filters, etc., all the way to a scenario where we have all 6 photometric points. For each of those scenarios we run the classifier for \emph{every single possible combination of filters}. This allows us to determine what the best filter combinations are for classifying planets by metallicity or other physical properties of interest. 

The metric we use to assess the success of an algorithm is the mean and standard deviation of the \emph{accuracy} after the k-fold cross validation test. It should be noted that there are many other metrics to evaluate a multi-class classification algorithm (e.g. logarithmic losses and confusion matrices). We address these after our initial, top-level, analysis of all the algorithms. 

\section{Results}\label{sec:results}
Before interpreting the results of the multivariate analysis, it is useful to assess the data using simpler statistical techniques in order to gain intuition for the full data set. We begin our results by presenting a correlation analysis to assess the sensitivity of the optical photometric bands to our physical planet properties of interest. Then, we present our color-color diagrams with the filter combinations expected to separate the planets into physically motivated groups (e.g. by metallicity). Lastly, we present the results of the multivariate analysis described in Section \ref{sec:stats}.
\begin{figure*}[ht]
\begin{center}
 \includegraphics[origin=c,width=0.6\linewidth]{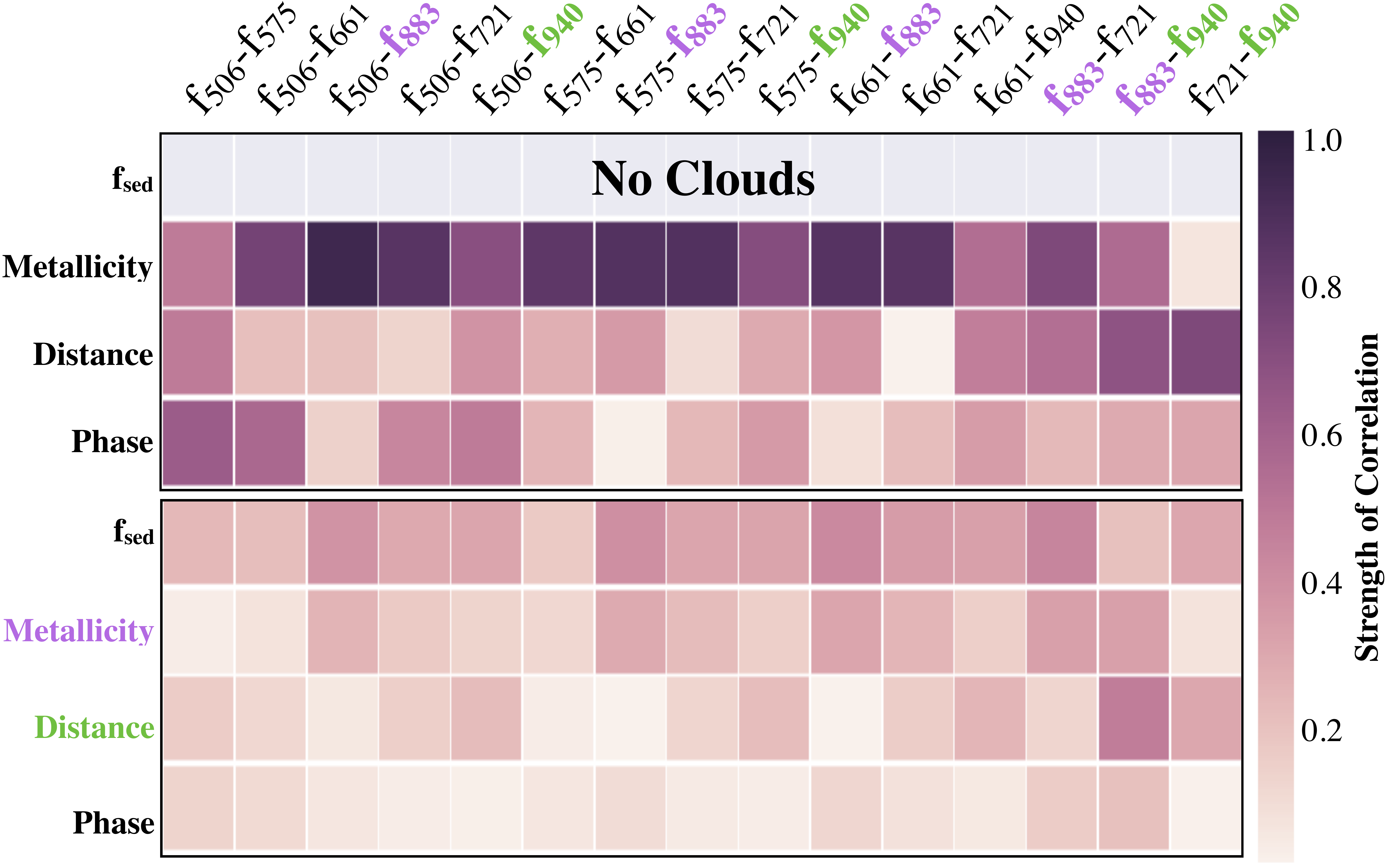}
\caption{Heatmap displays absolute correlations between retrievable parameters and differences in filters. Top panel shows strength of the correlations if no clouds were present in the sample of planets. Bottom panels shows the muted strength of correlations when clouds are added to the sample of planets. \textbf{Main point}: Optical photometric filters are only strongly correlated with planet properties, such as metallicity, when the effects of clouds are neglected. 
\label{fig:corrmat}}
\end{center}
\end{figure*}
\subsection{Sensitivity of Optical Photometric Bands to Planet Properties}
Figure \ref{fig:corrmat} shows the Pearson correlation matrix computed with Python's \texttt{pandas} package\footnote{\texttt{pandas.DataFrame.corr()}} \citep{mckinney2010data}. Correlation matrices are used to assess dependencies between variables. Therefore, it allows for the quick visualization of any dependencies between observational filters and physical planet properties. Filters where correlations exist, could be helpful in separating out planet populations in color-color space. Because we are purely interested in the relative strength of correlation with respect to the physical parameters of interest, we show the absolute value, for simplicity. We emphasize that this analysis is purely used to gain intuition for the full data set.

The top panel of the Figure \ref{fig:corrmat} shows the strength of correlations for a \emph{cloud-free} subset of the data. Metallicity exhibits strong correlations with filter combinations that include f$_{506}$ or f$_{575}$ minus f$_{721}$ or f$_{883}$. Distance/temperature exhibits the strongest correlations with filter combinations that include f$_{721}$, f$_{883}$, and f$_{940}$. It is also sensitive to  f$_{506}$-f$_{575}$ where the structure of the albedo spectrum is largely dictated by the presence or absence of water absorption. Phase also exhibits the strongest correlation to f$_{506}-$f$_{575}$. The presence of strong correlations in the \emph{cloud-free} subset of the data suggests that classifying planets given WFIRST-like filters might yield fruitful results. However, the bottom panel of Figure \ref{fig:corrmat} alludes to the main challenge we will encounter in classifying planets. 

The bottom panel of Figure \ref{fig:corrmat} shows the same Pearson correlation matrix with the full dataset (cloud-free plus cloudy). With the addition of $f_{\rm sed}$, all previous strong correlations become muted. Additionally $f_{\rm sed}$ now holds stronger correlations than metallicity between f$_{506}$ or f$_{575}$ minus f$_{721}$ or f$_{883}$. Therefore, we should expect any behavior in color-color diagrams to be primarily driven by cloud properties and only weakly driven by metallicty and temperature/distance. 
\begin{figure*}[ht]
\begin{center}
 \includegraphics[origin=c,width=\linewidth]{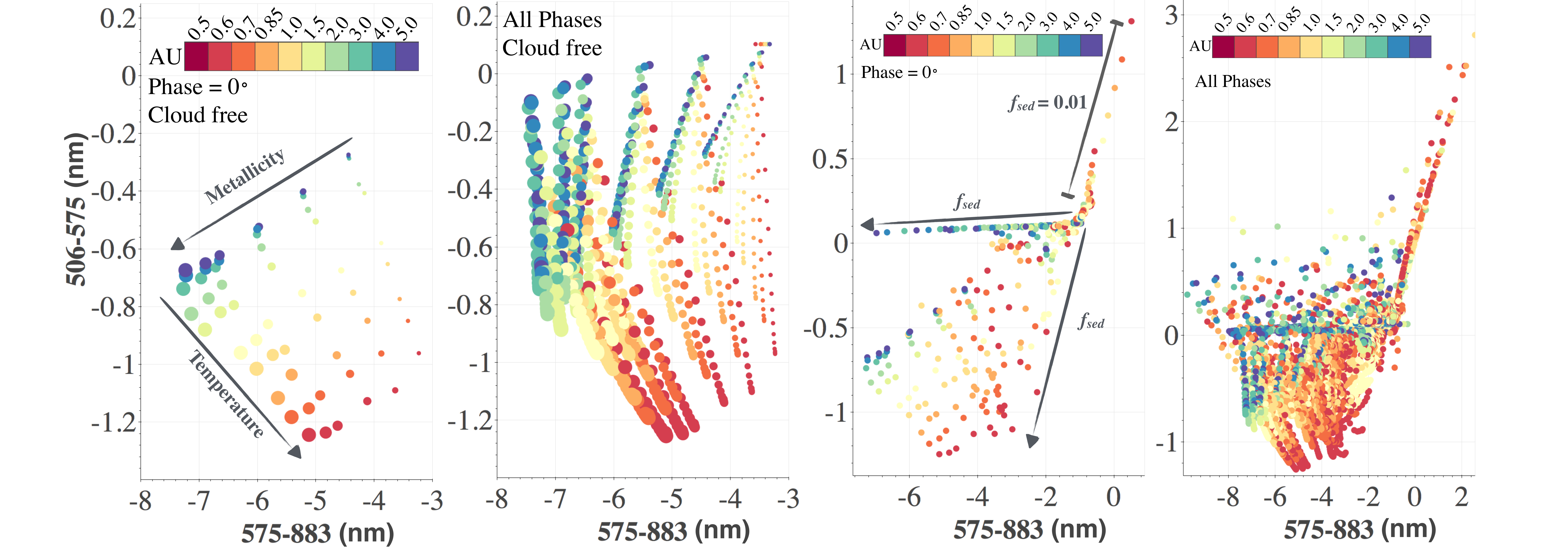}
\caption{Planet color-color plots with varying levels of complexity. From left to right, we first show a cloud free, sample with only geometric albedos. The symbol size increases with increasing metallicity. In this subset, planets can be visually separated by both temperature and metallicity. Next, we add all phase angles in a cloud-free sample to illustrate the increase in complexity. Even with the addition of phase, general metallicity and temperature/distance trends still remain. Finally, we add in all cloud models and return to zero phase (third from the left). Now the color-color space is confounded by effects of clouds. We show the approximate trend of $f_{\rm sed}$, the sedimentation efficiency, to illustrate the new structure of color-color diagram. The last plot shows the entire population of simulated planets.  \textbf{Main Point}: With a cloud-free sample, color-color diagrams for planets easily show trends in metallicity or temperature/distance. However, when the total cloudy and cloud-free sample is mixed, those trends are lost. 
\label{fig:cc_cf}}
\end{center}
\end{figure*}
\subsection{Color-Color Diagrams For Exoplanets}
Moving forward, we display our color-color diagrams with the combination of three filters: f$_{506}-$f$_{575}$ and f$_{575}-$f$_{883}$ in order to illustrate the general trends of color-color diagrams for planets. We choose f$_{883}$ because of its strong correlations with $f_{\rm sed}$, and metallicity, however we note that there are very few emergent photons at those wavelengths (see Figure \ref{fig:meh+au} at  $\sim$940 nm) and the SNR is expected to be relatively low \citep{lacy2018characterization}. Note that our full color grid is also available for download along with a visualization application to explore additional color-color parameter space and recreate Figure~\ref{fig:cc_cf}\footnote{\url{https://natashabatalha.github.io/color-color.html}}.

The leftmost panel in Figure \ref{fig:cc_cf} shows the cloud-free, zero phase, subset of our simulated grid of directly imaged planets. Temperature/distance is separated on the y-axis by f$_{506}-$f$_{575}$ because towards high temperatures, CH$_4$ does not dominate the structure of the albedo spectrum. Instead, Rayleigh scattering creates a deeper slope between f$_{506}$ and f$_{575}$. Metallicity, on the other hand, is separated on the x-axis by f$_{575}-$f$_{883}$ because the total atmospheric opacity darkens the planet in the f$_{575}$ bandpass at a faster rate than that at the f$_{883}$ bandpass. Overall, metallicity and temperature/distance are well separated in color-space, with the exception of planets with $a_s\ge3$~AU and [M/H]$\ge30\times$~Solar (an effect which is also seen in Figure \ref{fig:meh+au} and \citet{cahoy2010exoplanet}).

Adding varying phases (second from the left in Figure~\ref{fig:cc_cf}) adds an additional level of complexity. However, general trends in metallicity and temperature/distance still remain overall unchanged --- majority of low metallicity systems still have largest values of f$_{575}-$f$_{883}$ and majority of $a_s\le1$~AU still have the lowest values of f$_{506}-$f$_{575}$. 

Next, with the addition of all the cloud profiles (3rd plot from the left), the color-color plot takes on a new structure. The cloud-free sample is now degenerate with additional planets with large $f_{\rm sed}$ (thin cloud profiles). Additionally, there is a new population of planets with f$_{506}-$f$_{575}>0$. This theoretical population consists of models with $f_{\rm sed}$ values of $f_{\rm sed}=0.01$, and distances of $a_s\le$1~AU. Further investigation revealed these high temperature, low $f_{\rm sed}$ cases, produced vertically thick alkali clouds of Na$_2$S and ZnS with small particles. We discuss this behavior further in Section 5.1. 
\begin{figure*}[ht]
\begin{center}
 \includegraphics[origin=c,width=0.7\linewidth]{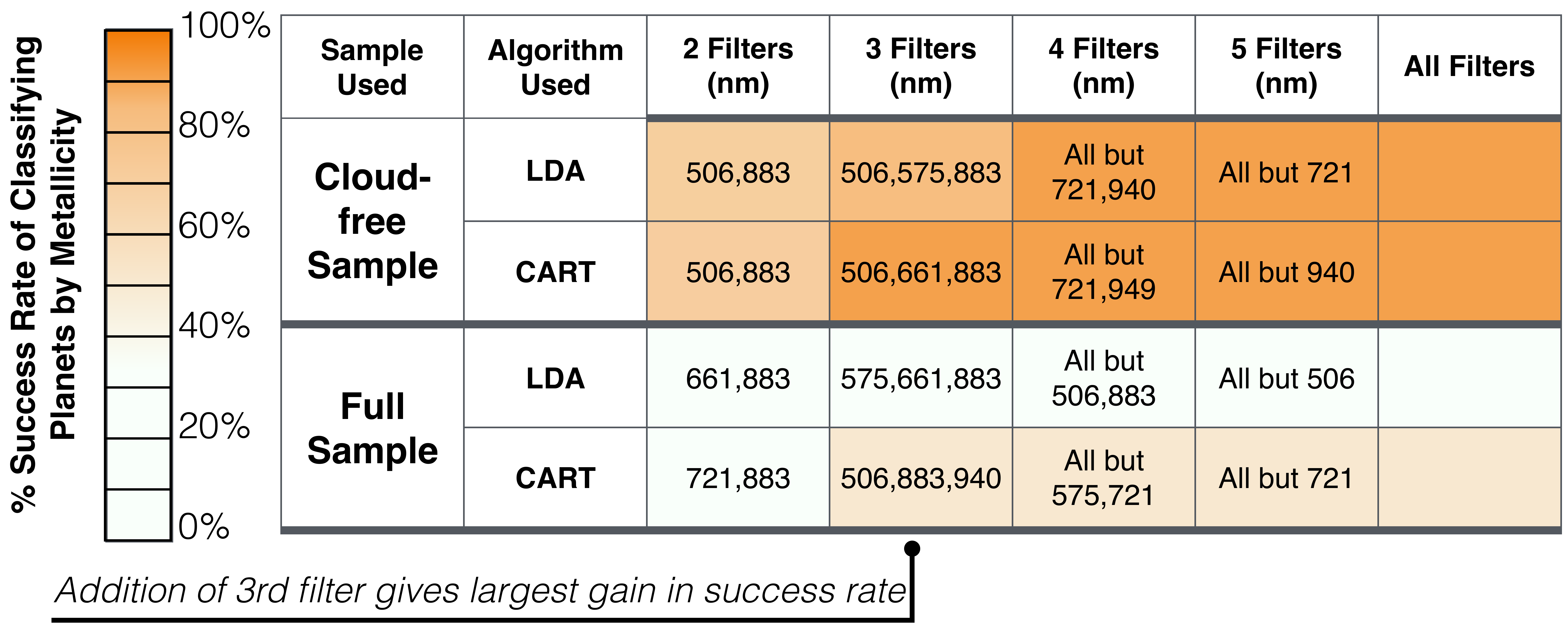}
\caption{The table summarizes the results of the two most successful algorithms out of the several that were tested: Linear Discriminant Analysis (LDA) and Classification and Regression Trees (CART). Cells are colored by the success rate (or \% accuracy) of the classifier. Each algorithm was run under different observing scenarios: observations in 2 filters, 3 filters, etc. The numbers in each cell are the optimized set of filter choices (nm) that yielded the maximum \% accuracy. \textbf{Main Point}: Classification algorithms yield high accuracy for a cloud free sample, only, with at least 3 filter observations. 
\label{fig:table}}
\end{center}
\end{figure*}
\begin{figure*}[ht]
\begin{center}
 \includegraphics[origin=c,width=\linewidth]{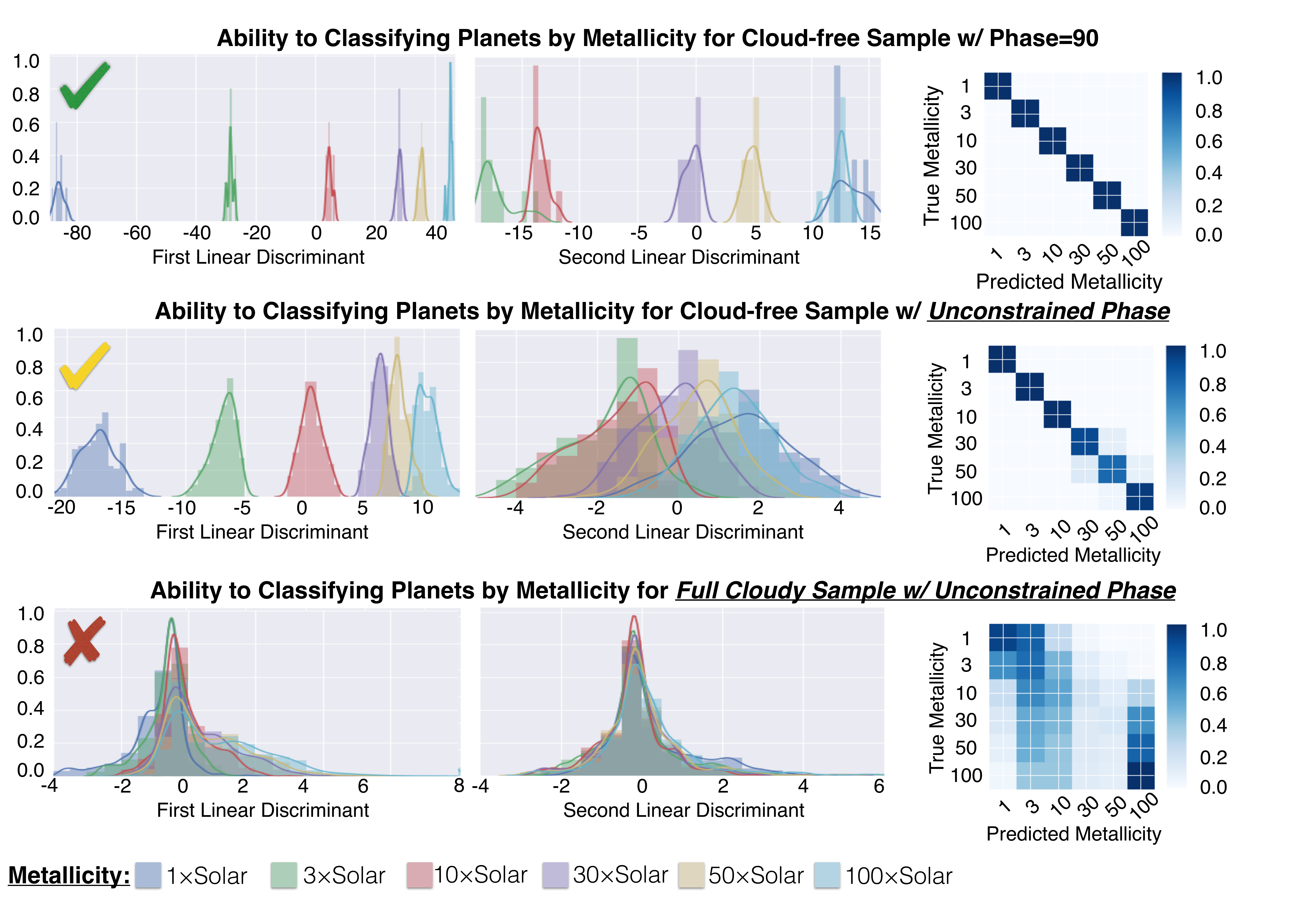}
\caption{Graphical representation of how planet classification breaks down using a Linear Discriminant Analysis. The first two columns of figures show the first and second Linear Discriminant (LD). A ``successfull'' classification is one in which the Linear Discriminants separate the planets into their respective classes. The LDA analysis succeeds with nearly 100\% accuracy when classifying a cloud-free planet at a single phase (top panel). When all possible phase angles are added, the first LD is unable to separate metallicities higher than 30$\times$~Solar (middle panel). When LDA is used on the full sample, it is unable to classify planets by their metallicity.  \textbf{Main Point:} Our ability to classify planet populations by metallicity diminishes when we have an unconstrained phase, and completely disappears with the presence of clouds. 
\label{fig:lda}}
\end{center}
\end{figure*}
\subsection{Classifying Planets by Colors: Where it works \& where it fails}
Our results in Section \ref{sec:results}.1 and \ref{sec:results}.2 indicate that although strong correlations exist between optical filters and physical planet properties, those correlations are visibly diminished with a large population of planets with variable cloud profiles. Here, we use the statistical algorithms described in Section \ref{sec:stats} to quantitatively assess whether or not this remains true. 

We first test the algorithms on a purely cloud-free sample to determine whether or not they succeed in the simplest of cases. Our results for the LDA and CART analyses are summarized in Figure \ref{fig:table}. We only show results for the algorithms that had mean accuracy of greater than 50\%. SVM had the poorest performance in the cloud-free sample, with less than 1\% accuracy. KNN reached a highest mean accuracy of 41\% in the case where all 6 filter combinations were available. LDA and CART  both succeeded in classifying planets by metallicity with greater than 90\% accuracy when a cloud-free sample was used. Additionally, they both gained the greatest accuracy with the addition of a third filter. Four filters did increase the accuracy of LDA, but not CART. The difference in performance of each algorithm depends largely on the non-linearity of the problem (e.g. whether an increase in metallicity is always linearity correlated with a photometric point). Some classifiers, such as SVM and logistic regression, depend more heavily on linear features in the data as opposed to CART, for example. CART has the disadvantage that it is susceptible to over-fitting the data. We take care to ensure our analysis is incentive to this by scaling back the maximum levels allowed in the decision tree, discussed below.  

Figure \ref{fig:table} also shows the optimized filter selection of each case study. For the cloud-free case, both algorithms agree that the filters at 506~nm and 883~nm are highest priority for planet classification. It is interesting to note that they do not agree on the third highest priority filter. LDA prefers the filter at 575 nm, while CART prefers the filter at 661 nm. However, both algorithms agree that the top four highest priority filters for planet classification in a cloud-free sample are the filters at: 506~nm, 883~nm, 575~nm, and 661~nm (approximately in that order). These results intuitively agree with the results of Figure \ref{fig:corrmat}.

Upon introducing the full sample, we first tune/retrain the model. Figure \ref{fig:table} again, shows the results of the LDA and CART. SVM still performs the poorest with less than 1\% accuracy. KNN again only reaches an accuracy of 40\% when all 6 photometric points are used. LDA, which was successful before, now can only classify planets by metallicity with $\sim$20\% accuracy. CART has the highest performance out of all of the algorithms, however still only attains a maximum success rate of 55\%. This analysis suggests that classifying planets via photometric analysis of colors is incredibly difficult due to the variety of cloud compositions. 

We also explore our ability to classify by metallicity if the cloud-free population is excluded from the population entirely. This is motivated by the prevalence clouds in exoplanet \citep[e.g.][]{sing2016continuum} and giant Solar System planet observations \citep[e.g.][]{lewis1969clouds,baines1995clouds}. However, limiting ourselves to just the giant planets that have clouds, we only can attain a classification accuracy of 45\%, suggesting that the prevalence of clouds inhibits our ability to accurately classify by metallicity with color observations alone.

Figure \ref{fig:lda} graphically shows how and when the classifications scheme breaks down. The top panel shows our population of planets separated by the linear discriminants computed in the LDA for the cloud-free sample of planets at a single phase. The first linear discriminant (LD) is successful in separating all metallicities. The heat map to the right shows the confusion matrix, another metric for evaluating machine learning algorithms. The confusion matrix dictates, for each class, what fraction of planets we have correctly predicted. In this cloud-free, Phase=90$^\circ$ sample we have correctly predicted nearly 100\% of cases. Next we add the full sample of phases and rerun the LDA. The first LD now has trouble separating metallicities greater than 30$\times$ Solar and the second LD cannot further help distinguish between those cases. This means the predictive power of the algorithm is decreasing. With the addition of the full cloudy sample the first LD and the second LD fail to separate planets by metallicity.

Next, we repeat the analysis above in order to group planets by cloud profile ($f_{\rm sed}$) and planet-star separation. We are unable to classify by planet-separation with higher than 47\% accuracy. For cloud profile, with the full sample of planets, CART is able to classify planets by $f_{\rm sed}$ with 83\% accuracy with 3 filters. Our optimization of the CART analysis yielded a 16-level decision tree. Decreasing the maximum depth to 5 yields a lower accuracy of 75\%, yet this is still a much higher performance than our ability to classify populations based on metallicity. Lastly, using a different criterion for distinguishing impurities (information gain versus Gini impurity) still yields an accuracy of 80\% in the full sample. In order to yield this accuracy, observations must be taken in at least the band passes at 575~nm, 883~nm, and 940~nm. 

Lastly, in order to verify that these results are insensitive to filter selection, we adopt the ``optimized'' filters from \citet{krissansen2016is}. With the new set of filters our results do not change. That is, we are 1) still unable to classify our directly imaged planets by metallicity unless we specifically select a cloud-free population, 2) able to classify the full sample by clouds with an accuracy of 75\% with CART, and 3) unable to classify by planet-star separation/temperature. 

\subsubsection{Binary Classification: Cloudy or not?}
The results depicted in Figure \ref{fig:table} \& \ref{fig:lda} suggest that determining accurate physical parameters such as metallicity, cloud composition, temperature, or phase, is unattainable due to the complex structures that diverse cloud species introduce to the full sample. However, photometry could still be potentially identifying targets for more time intensive follow-up. As Figure \ref{fig:cloud} shows, very cloudy cases exhibit little or no spectral features, which would not be ideal candidates for follow up. In this case, classifying giant planets based on whether they are cloudy or cloud-free would be very useful. We find that a binary supervised classification attains a maximum accuracy of 97\% with 3 filters. Therefore, photometric observations will valuable for identifying targets for more intensive spectroscopic follow-up. 

\section{Discussion}\label{sec:disc}
\begin{figure}[ht]
\begin{center}
 \includegraphics[origin=c,width=\linewidth]{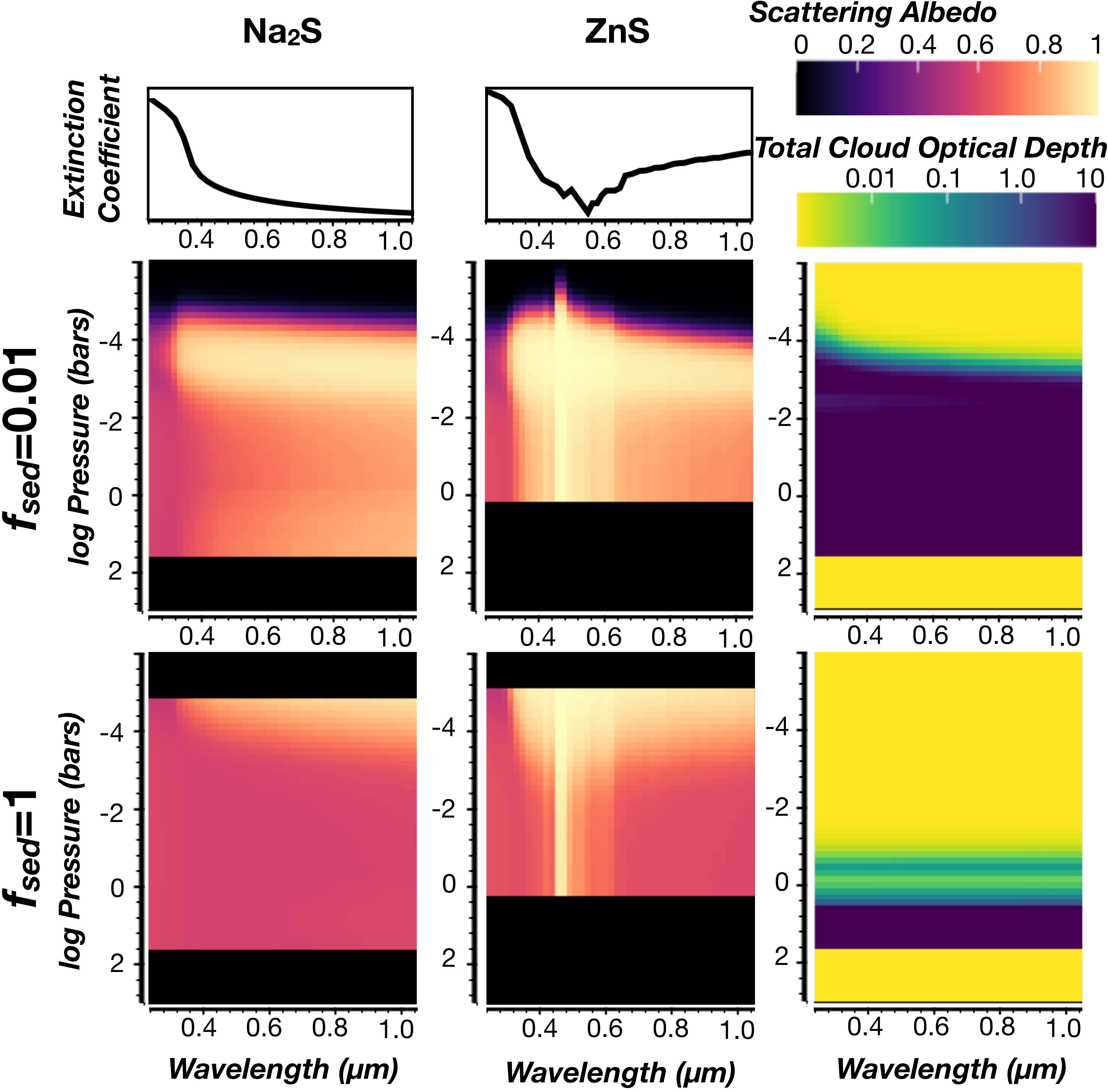}
\caption{Vertical- and wavelength-dependent cloud profiles for Na$_2$S and ZnS at different values of f$_{sed}$. The top row shows the extinction coefficient for each cloud species, used to describe their absorbing properties. \textbf{Main point:} Hot temperature clouds of Na$_2$S and ZnS strongly absorb short of 0.6 $\mu$m. At low f$_{sed}$=0.01, where small particles create vertically thick clouds, this results in spectra with suppressed blue light (see the the arm in Figure \ref{fig:cc_cf}). 
\label{fig:alkali}}
\end{center}
\end{figure}

\begin{figure}[ht]
\begin{center}
 \includegraphics[origin=c,width=\linewidth]{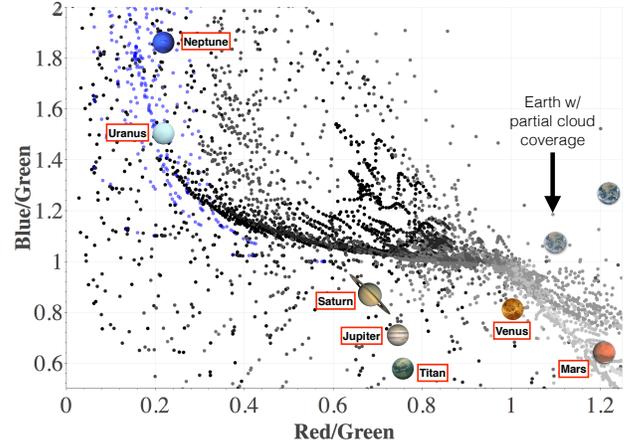}
\caption{The full grid of directly imaged planets as seen in the ``optimized'' filter selection proposed by \citet{krissansen2016is}, compared to the Solar System planets. \textbf{Main Point}: ``Optimized'' filter selection does not separate Solar System planets and theoretical population into physically motivated groups. 
\label{fig:ss}}
\end{center}
\end{figure}
\subsection{Population of planets at low $f_{\rm sed}$}
Figure \ref{fig:cc_cf} shows a unique population of theoretical planets with very bright red (883 nm) components, as compared to the blue (506 nm). Here we further explore that population. The population consists of theoretical planets with very low sedimentation efficiency ($f_{\rm sed}=0.01$) and high temperatures ($a_s<1$~AU). Figure \ref{fig:alkali} shows the wavelength- and vertical-dependent cloud profiles for the alkali clouds (Na$_2$S and ZnS) that form in this region of parameter space. At $f_{\rm sed}=1$, the clouds become optically thick at high pressures (1~bar) and are vertically thin (only 2 dex in pressure). However, at very low $f_{\rm sed}=0.01$ the particles become small, and the clouds become vertically extended, more closely simulating the effect of hazes. Figure \ref{fig:alkali} also shows the extinction coefficient of Na$_2$S \citep{montaner197na2s, khachai2009na2s} and ZnS \citep{querry1987optical}. The extinction coefficient, also referred to as the imaginary part of the index of refraction, represent the absorbing power of each cloud species. The absorbing power of sulfur alkali species strongly suppress the blue part of the reflected light spectra, a result that has also been seen in studies of sulfur produced hazes \citep{gao2017sulfur}.

Thus far, low values of $f_{\rm sed}$ have only been observed in close-in hot Jupiters \citep{demory2013inference}. Observations of Jupiter in reflected light suggest that widely separated exoplanets might be limited to $f_{\rm sed} \ge 3$. Therefore, we additionally explore the feasibility of classifying a sample that only consists of $f_{\rm sed} \ge 3$. We find that classifying by metallicity and by cloud with 3 imaging filters still results in a maximum accuracy of 48\% and 62\%, respectively. The accuracy does not increase when the sample of $f_{\rm sed}$ is confined because small spectral changes introduced by vertically thin clouds are photometrically degenerate with changes in phase, temperature, or metallicity. See also \citet{macdonald2018exploring} for an exploration of how spectral morphology depends upon $f_{\rm sed}$.

\subsection{Where does the Solar System Fit in the Population?}
One question not yet addressed is where the Solar System fits within our giant planet population. Figure \ref{fig:ss} shows the Solar System planets in the color space chosen by \citet{krissansen2016is} to separate biogenic Earth from lifeless planets. Their filter choice is specifically sensitive to the ``U-shape'' created by Rayleigh scattering and ozone absorption. Disregarding our population of giant planets, the Solar System planets do appear to exist in unique color spaces. The ice giants, Neptune and Uranus are brighter in the blue, and darker in the red, as compared to Jupiter, Saturn and Titan. The terrestrial planets are bright in the red, with Earth having an extra blue component from its unique ``U-shape''. However, when the additional population of giant planets is considered, the unique separation of these planets is washed out. Many of our giant planet cases with $f_{\rm sed}\le1$ are easily degenerate in color space to Mars and Venus. While there are significantly fewer ``false-positives'' near Earth, a more realistic partial cloud coverage model of Earth \citep[from][]{krissansen2016is} is also synonymous in color space to many of our giant planet models. 

Specifically, at $f_{\rm sed}\sim0.3$ and at $d<1$~AU: 1) temperatures are high enough so as to avoid formation of bright water clouds, 2) sedimentation efficiency is low enough to obscure alkali opacity features, and 3) sedimentation efficiency is high enough to retain the downward slope Rayleigh opacity. These three features result in the classic ``Earth-like'' Rayleigh slope, but of course lack a large upward slope from ozone absorption. However, when ozone is partially obscured by clouds (i.e Earth w/ partial cloud coverage in Figure \ref{fig:ss}), low resolution photometric observations could result in degeneracies. Additional intricacies could further confound this parameter space.

\subsection{Additional Intricacies: Unknown Gravity, Potential Hazes, \& Composition}
The three parameters not explored in this work are the effect of poorly unconstrained or unknown gravities \citep{nayak2017atmospheric,lacy2018characterization}, potential hazes \citep{gao2017sulfur}, and unknown chemical compositions (e.g. non-solar C/O ratio). 

Increasing radii is degenerate with decreasing phase \citep[see Fig. 1][]{nayak2017atmospheric}. Gravity alone also impacts the overall structure of the reflected light spectrum \citep[see Fig. 5][]{macdonald2018exploring}. Both \citet{nayak2017atmospheric} and \citep{lacy2018characterization} were able to constrain radii to relatively high fidelity with spectroscopy. This is partly because at higher (more crescent) phases, the atmospheres becomes forward scattering, and the effect of clouds become less dominant. However, photometry will be much less forgiving, even with three or four filter observations. Adding additional gravities to our grid will ultimately decrease the accuracy of our classification algorithms. 

\citet{gao2017sulfur} investigated the effects of sulfur hazes in giant exoplanet atmospheres in reflected light and found that haze masses drastically alter a planet's geometric albedo spectrum. These sulfur hazes mimic the behavior of our very low $f_{\rm sed}<0.03$ cases with sulfur-based alkali clouds (see Figure \ref{fig:alkali}. Therefore, accounting for additional hazes would not drastically change the accuracy of our classification algorithm when classifying by cloud. Instead they would be classified in the same grouping as our low $f_{\rm sed}$ cases.

Lastly, compositions that deviate significantly from solar C/O would also change the accuracy of our classifiers if new molecular features became dominant. At the temperatures explored in this work, the chemically favorable molecule, regardless of C/O is expected to be CH$_4$ and H$_2$O \citep{madhu2012co}. The atmospheric $\rm H_2O$ abundance has not yet been measured on Jupiter, so detailed studies of this parameter are premature. Nevertheless, we would not expect varying C/O to largely change the results presented here.

\subsection{Prospects for Filter Optimization Techniques}
Although these additional sources of uncertainty have the potential to degrade our ability to classify planets, there are further steps we can take to improve our ability to accurately classify planets in color space. There is currently a large  parameter space to explore in filter choices. The \emph{WFIRST} filters have not yet been solidified. Therefore, our pipeline can be used determine the optimal filter design in the era of \emph{WFIRST} and future mission concepts. 

In order to test the feasibility of using our pipeline for this type of analysis, we determined whether or not there was any 5\% bandpass filter from 0.5-1~$\mu$m that could be added to a 575~nm photometric point to greatly increase classification accuracy. This was specifically chosen because the \emph{WFIRST} science investigation teams are currently considering a 575~nm engineering filter in conjunction with just one more commissioned filter. With just two filters, we did not find any combination of 575~nm+a second filter that greatly improved the prospects of classifying our giant planet population. Regardless of the additional filter from 0.5-1$\mu$m, we achieved a classification accuracy of just 25\% when classifying by metallicity and just 50\% when classifying by $f_{\rm sed}$. 

This result generally agrees with  Figure \ref{fig:table}, which shows that the addition of a third filter is needed to attain a relatively high classification accuracy. This simple parameter space analysis reiterates the importance of having at least three filter observations. Additionally, it is possible that decreasing the bandpass (from 5\%) and increasing the wavelength range to explore would yield a more favorable optimized filter selection. We leave a comprehensive analysis of these intricate choices and the additional uncertainties described above for a future analysis.

\section{Conclusion}\label{sec:conc}
By analyzing a grid of 9,120 theoretical gas-giant reflected light spectra, we explored our ability to classify giant planets into physically motivated populations. Specifically, we were interested in determining if the proposed \emph{WFIRST} photometric bins could classify giant planets by either their temperature, metallicity or cloud properties. In order to do so we used several multivariate classification algorithms. Out of the six algorithms tested, only the Linear Discriminant Analysis (LDA) and Classification and Regression Trees (CART) yielded fruitful results. However, neither algorithm was able to yield very high accuracy when classifying by physical planet parameters. 

In accordance with the questions posted in Section 1, we found that: 
\begin{enumerate}
    \item There is a strong correlation between atmospheric properties and \emph{WFIRST}-like optical filters (Fig. \ref{fig:corrmat}). However, these correlations only truly exist for a population that does not have significant cloud coverage in the visible part of the atmosphere. If a full sample of cloud-free and cloudy planets is considered, there are less strong correlations between atmospheric properties and photometric bins that can be leveraged to classify planets. 
    \item For giant panets, it is only possible to classify planets into physically motivated groups with greater than 90\% accuracy if it is known \emph{a priori} that the planet does not have significant cloud coverage  in the visible. However, observations of Solar System and exoplanet giant planets suggest clouds are prevalent in most planetary atmospheres.  
    \item Our machine learning algorithms are unable to classify planets by metallicity with greater than 55\% accuracy. However, we are able to classify planets with moderate accuracy $\sim$70\% by classifying by the cloud sedimentation efficiency, $f_{\rm sed}$. Additionally, binary classification between cloudy and cloud-free populations are successful with an accuracy $>$90\%. 
    \item We find that at least three filters are needed for any kind of classification (cloud, metallicity, etc). We also tested our classification algorithm with another filter set proposed by \citet{krissansen2016is}, but do not find more optimistic results. However, our statistical algorithm is open-source and can be used to determine optimal filters for \emph{WFIRST} or future mission concepts. 
\end{enumerate}

These results highlight the difficulties with classifying giant extrasolar planets with photometry alone. However, photometry will still be a useful tool for planet searches and potentially identifying targets for follow-up. 

\acknowledgments

We thank STScI's STARGATE team for helpful comments and Caroline Morley for critical guidance in running theoretical models. The authors acknowledge support from the NASA Grant \# NNG16PJ24C and Grant \# NNX15AJ80G.

\software{scikit-learn \citep{scikit-learn}, pandas \citep{mckinney2010data}, bokeh \citep{bokeh}, NumPy \citep{walt2011numpy}, IPython \citep{perez2007ipython}, Jupyter, \citep{kluyver2016jupyter}, SciPy \citep{jones2014scipy}}

\bibliography{apjmnemonic,albedo_wfirst}
\bibliographystyle{apj}

\end{document}